\documentstyle[aps]{revtex}
\begin{document}
\title{Multi-output programmable quantum processor}
\author{Yafei Yu$^{\thanks{%
corresponding author, e-mail: yfyu@wipm.ac.cn}a}$ , Jian Feng$^{a,b}$ ,
Mingsheng Zhan$^{a}$}
\address{$^{a}$State Key Laboratory of Magnetic Resonance and Atomic and Molecular\\
Physics,\\
Wuhan Institute of Physics and Mathematics, Chinese Academy of Sciences,\\
Wuhan 430071, PR China\\
$^{b}$Institute of Optical Communication, Liaocheng University, Liaocheng ,\\
252059, Shandong, PR China}
\maketitle

\begin{abstract}
By combining telecloning and programmable quantum gate array presented by
Nielsen and Chuang [Phys.Rev.Lett. 79 :321(1997)], we propose a programmable
quantum processor which can be programmed to implement restricted set of
operations with several identical data outputs. The outputs are
approximately-transformed versions of input data. The processor successes
with certain probability.

PACS number(s): 03.67.Lx
\end{abstract}

\section{Introduction}

Quantum processor stems from analogy between quantum computer and classical
computer. Similarly as in classical computer a fixed classical gate array,
which takes the program and data as input, performs operations on the data
depending on the program, quantum processor consists of a fixed quantum gate
array, a quantum system as data register carrying data state and another
quantum system as program register containing information about operation to
be performed on the data state. When the data state and program state go
through the fixed quantum gate array together, an unitary operation is
executed on the data state depending on the program state. But it is shown
by Nielsen and Chuang \cite{nielsen and chuang} that in quantum computer no
such fixed quantum gate array can be realized in deterministic fashion.
Suppose that the action of a fixed quantum gate array is operator $G$, $%
\left| d\right\rangle $ is a data state and $\left| P_{U}\right\rangle $ is
a program state, the total dynamics of the quantum processor can be
described as \cite{nielsen and chuang}:

\begin{equation}
G(\left| d\right\rangle \otimes \left| P_{U}\right\rangle )=\left( U\left|
d\right\rangle \right) \otimes \left| P_{U}^{^{\prime }}\right\rangle 
\end{equation}
where $U$ is a unitary operation corresponding to the state $\left|
P_{U}\right\rangle $. From Eq. (1) it follows that distinct (up to a global
phase ) unitary operation $U$ requires different orthogonal state of program
system. It means that the dimension of the program system has to be infinite
because even for a single qubit the set of operations is infinite. Therefore
no universal quantum gate array can be constructed in deterministic fashion.
Fortunately there is possibility to build a universal quantum gate array in
a probabilistic fashion. An example based on teleportation is given in Ref.%
\cite{nielsen and chuang}. Another explanation in terms of control-U
operation is given to realize a one-qubit arbitrary rotation around the $z$
axis of a spin $\frac{1}{2}$ particle, in which the probability of success
increases exponentially with the number of qubits of the program register 
\cite{vidal and cirac}. On the other hand it is possible to construct a
programmable quantum gate to implement an interesting subclass of unitary
operations perfectly, accordingly it is possible to implement an arbitrary
operation probabilisticly which can be decomposed into linear combinations
of the subclass of unitary operations. The problem is recently examined by
Hillery et. al. and generalized to the case of qudits of dimension $N>2$ 
\cite{hillery and ziman}. Also a mathematical description of programmable
quantum processor is developed and its classes are analyzed by the same group%
\cite{hillery and buzek}. We note that the processors in the previous works 
\cite{nielsen and chuang,vidal and cirac} are from an input to an output. In
quantum network one-input-to-multiple-output processor is more desirable. So
it is worthwhile to extend the one-to-one processor to the case of multiple
outputs.

In this paper we combine telecloning \cite{telecloning} and programmable
quantum gate array based on teleportation \cite{nielsen and chuang} to
discuss the possibility of multi-output quantum processor for restricted
sets of operations commuting and anti-commuting with Pauli operator $\sigma
_{z}$. Here we encode operation into a particular multiparticle-entangled
state instead of maximally two-qubit entangled state. In addition, we
slightly modify the procedure of the quantum gate array and redefine the
location of the data register output. Similarly, when joint Bell-state
measurement gives some appropriate values, several identical states come out
in the data register outputs. The output data states are
approximately-transformed versions of input data state. Then these several
approximate data outputs could be distributed to next different quantum
processors for different tasks as described in \cite{1022301}. Though,
compared with one-to-one processors in Refs.\cite{nielsen and chuang,vidal
and cirac}, our multi-output processor has several disadvantages, it has
extensive applications in remote control and distributed computation. For
simplicity, we focus on single qubit operation and two data outputs in the
following discussion. We elaborate the details in Sec.II, and give our
discussion in Sec.III and conclusion in Sec.IV.

\section{Two-output quantum gate array}

In this section,we will discuss a two-output quantum processor which
produces two data register outputs for sets of operations on single qubit
commuting or anti-commuting with Pauli operator $\sigma _{z}$. In the
processor the operation is encoded into a four-qubit entangled state and the
quantum gate array is similar with one proposed by Nielsen and Chuang \cite
{nielsen and chuang}. After the Bell-state measurement, two approximate data
register outputs appear at the terminal with probability of $\frac{1}{2}$,
and the fidelity of each data register output is $\frac{5}{6}$.

First we consider how the quantum gate array implement the operation
commuting with operator $\sigma _{z}$with two data outputs. The set of
operations has the form of 
\begin{equation}
U=\exp (i\alpha \frac{\sigma _{z}}{2})=\left( 
\begin{array}{cc}
e^{i\frac{\alpha }{2}} & 0 \\ 
0 & e^{-i\frac{\alpha }{2}}
\end{array}
\right) ,
\end{equation}
where an arbitrary angle $\alpha $ $\in [0,2\pi )$. The gate array is
illustrated in Fig.1.

The four-qubit entangled state $\left| \Psi \right\rangle $ to be codified
into program state is formulated as: 
\begin{equation}
\left| \Psi \right\rangle =\frac{1}{\sqrt{2}}(\left| 0\right\rangle
_{P}\left| \Phi _{0}\right\rangle _{ABC}+\left| 1\right\rangle _{P}\left|
\Phi _{1}\right\rangle _{ABC})
\end{equation}
where qubit $P$ is a port qubit subjected to Bell-state measurement M with
the qubit as data input register, qubits A, B and C serve as data output
register together, qubit A is an ancillary which is discarded eventually.
The sates $\left| \Phi _{0}\right\rangle _{ABC}$ and $\left| \Phi
_{1}\right\rangle _{ABC}$ are given, respectively as: 
\begin{equation}
\left| \Phi _{0}\right\rangle _{ABC}=\sqrt{\frac{1}{3}}\left|
000\right\rangle +\sqrt{\frac{1}{6}}(\left| 101\right\rangle +\left|
110\right\rangle )
\end{equation}
\begin{equation}
\left| \Phi _{1}\right\rangle _{ABC}=\sqrt{\frac{1}{3}}\left|
111\right\rangle +\sqrt{\frac{1}{6}}(\left| 001\right\rangle +\left|
010\right\rangle )
\end{equation}
where the subscript is neglected for concision, the qubits are arranged in
accordance with the order A, B, C from the left to the right.

In order to prepare a program state, the state $\left| \Psi \right\rangle $
is transformed through $U_{P}\otimes I_{ABC}$, that is, 
\begin{equation}
\left| P_{U}\right\rangle =U_{P}\otimes I_{ABC}\left| \Psi \right\rangle =%
\frac{1}{\sqrt{2}}(e^{i\frac{\alpha }{2}}\left| 0\right\rangle \left| \Phi
_{0}\right\rangle +e^{-i\frac{\alpha }{2}}\left| 1\right\rangle \left| \Phi
_{1}\right\rangle )
\end{equation}
where only the port qubit P is altered by $U$. For an arbitrary input data
state $\left| d\right\rangle =a\left| 0\right\rangle +b\left| 1\right\rangle 
$, where $a$ and $b$ are complex numbers and satisfy $\left| a\right|
^{2}+\left| b\right| ^{2}=1$, the total state $\left| d\right\rangle \otimes
\left| P_{U}\right\rangle $ feeding into the gate array can be rewritten as: 
\begin{eqnarray}
\left| d\right\rangle \otimes \left| P_{U}\right\rangle &=&(a\left|
0\right\rangle +b\left| 1\right\rangle )\frac{1}{\sqrt{2}}(e^{i\frac{\alpha 
}{2}}\left| 0\right\rangle \left| \Phi _{0}\right\rangle +e^{-i\frac{\alpha 
}{2}}\left| 1\right\rangle \left| \Phi _{1}\right\rangle ) \\
&=&\frac{1}{2}[a(\left| \Phi ^{+}\right\rangle +\left| \Phi
^{-}\right\rangle )e^{i\frac{\alpha }{2}}\left| \Phi _{0}\right\rangle
+a(\left| \Psi ^{+}\right\rangle +\left| \Psi ^{-}\right\rangle )e^{-i\frac{%
\alpha }{2}}\left| \Phi _{1}\right\rangle  \nonumber \\
&&+b(\left| \Phi ^{+}\right\rangle -\left| \Phi ^{-}\right\rangle )e^{i\frac{%
\alpha }{2}}\left| \Phi _{0}\right\rangle +b(\left| \Psi ^{+}\right\rangle
-\left| \Psi ^{-}\right\rangle )e^{-i\frac{\alpha }{2}}\left| \Phi
_{1}\right\rangle ]  \nonumber \\
&=&\frac{1}{2}[\left| \Phi ^{+}\right\rangle (ae^{i\frac{\alpha }{2}}\left|
\Phi _{0}\right\rangle +be^{-i\frac{\alpha }{2}}\left| \Phi
_{1}\right\rangle )+\left| \Phi ^{-}\right\rangle (ae^{i\frac{\alpha }{2}%
}\left| \Phi _{0}\right\rangle -be^{-i\frac{\alpha }{2}}\left| \Phi
_{1}\right\rangle )  \nonumber \\
&&+\left| \Psi ^{+}\right\rangle (be^{i\frac{\alpha }{2}}\left| \Phi
_{0}\right\rangle +ae^{-i\frac{\alpha }{2}}\left| \Phi _{1}\right\rangle
)+\left| \Psi ^{-}\right\rangle (be^{i\frac{\alpha }{2}}\left| \Phi
_{0}\right\rangle -ae^{-i\frac{\alpha }{2}}\left| \Phi _{1}\right\rangle ), 
\nonumber
\end{eqnarray}
where $\left| \Psi ^{\pm }\right\rangle $ and $\left| \Phi ^{\pm
}\right\rangle $ are Bell states. When the measurement result from M gives
an eigenvalue corresponding to $\left| \Phi ^{+}\right\rangle $, the three
data qubits are projected onto a state 
\[
\left| X\right\rangle =(ae^{i\frac{\alpha }{2}}\left| \Phi _{0}\right\rangle
+be^{-i\frac{\alpha }{2}}\left| \Phi _{1}\right\rangle ). 
\]
Dropping the ancillary qubit A and calculating the reduced density matrices
on the qubits B and C respectively, we get 
\begin{eqnarray}
\rho _{_{B}} &=&Tr_{A,C}(\left| X\right\rangle \left\langle X\right| )=\rho
_{_{C}}=Tr_{A,B}(\left| X\right\rangle \left\langle X\right| ) \\
&=&\frac{5\left| a\right| ^{2}+\left| b\right| ^{2}}{6}\left| 0\right\rangle
\left\langle 0\right| +\frac{5\left| b\right| ^{2}+\left| a\right| ^{2}}{6}%
\left| 1\right\rangle \left\langle 1\right|  \nonumber \\
&&+\frac{2}{3}ab^{*}e^{i\alpha }\left| 0\right\rangle \left\langle 1\right| +%
\frac{2}{3}a^{*}be^{-i\alpha }\left| 1\right\rangle \left\langle 0\right| 
\nonumber
\end{eqnarray}
However the correctly transformed version of the input data state should be $%
U\left| d\right\rangle =ae^{i\frac{\alpha }{2}}\left| 0\right\rangle +be^{-i%
\frac{\alpha }{2}}\left| 1\right\rangle $. We estimate the fidelity between
the correct version and the resulting state as 
\begin{eqnarray}
F_{B} &=&\left\langle d\right| U^{\dagger }\rho _{out}U\left| d\right\rangle
\\
&=&(a^{*}e^{-i\frac{\alpha }{2}}\left\langle 0\right| +b^{*}e^{i\frac{\alpha 
}{2}}\left\langle 1\right| )\rho _{_{B}}(ae^{i\frac{\alpha }{2}}\left|
0\right\rangle +be^{-i\frac{\alpha }{2}}\left| 1\right\rangle )  \nonumber \\
&=&\frac{5}{6},  \nonumber \\
F_{C} &=&\frac{5}{6}.
\end{eqnarray}
Hence we obtain two data register outputs with fidelity of $\frac{5}{6}$.
When the measurement result from M gives an eigenvalue corresponding to $%
\left| \Phi ^{-}\right\rangle $, the three data qubits are projected onto a
state 
\begin{equation}
\left| Y\right\rangle =(ae^{i\frac{\alpha }{2}}\left| \Phi _{0}\right\rangle
-be^{-i\frac{\alpha }{2}}\left| \Phi _{1}\right\rangle ),
\end{equation}
which can be converted into the state $\left| X\right\rangle $ by applying
the operation $\sigma _{z}$ on all three data qubits. Again we can have the
resulting state in the fidelity of $\frac{5}{6}$ with the correctly
transformed version of data input state. For the two possible results of the
measurement M corresponding to $\left| \Psi ^{+}\right\rangle $ and $\left|
\Psi ^{-}\right\rangle $, the quantum gate array produces ruined state from
which the state $\left| X\right\rangle $ can not be retrieved effectively
and the process of quantum gate array on the input data state fails.
Consequently, from the results of measurement M we know whether the process
is successful. The four results of the measurement M are uniformly
distributed at random, containing no information about $U$ and $\left|
d\right\rangle $. So our processor is no deterministic and succeeds with
certain probability of $\frac{1}{2}$. Because in the process there is only
one data input but there are two data outputs, it implies a cloning
transformation. In fact, it involves a 1$\rightarrow $2 telecloning
procedure. Each output data state has a fidelity of $\frac{5}{6}$ with the
correctly transformed version of input data state. Thus our processor
consisting of a fixed gate array illustrated in Fig.1 and a particular
program register and a data register can implement arbitrary rotation around 
$z$ axis with two data outputs with fidelity of $\frac{5}{6}$ and successful
probability of $\frac{1}{2}$.

Next, we adjust our processor to implement operations of anti-commuting with
Pauli operator $\sigma _{z}$ with two data outputs. The set of operations
has the form of 
\begin{equation}
U^{^{\prime }}=\left( 
\begin{array}{cc}
0 & e^{i\frac{\beta }{2}} \\ 
e^{-i\frac{\beta }{2}} & 0
\end{array}
\right) ,
\end{equation}
where an arbitrary angle $\beta $ $\in [0,2\pi )$. The quantum gate array
modified is illustrated in Fig.2. The initial state of the program register
is reset as 
\begin{equation}
\left| \Phi \right\rangle =\frac{1}{\sqrt{2}}(\left| 0\right\rangle
_{P}\left| \Phi _{1}\right\rangle _{ABC}+\left| 1\right\rangle _{P}\left|
\Phi _{0}\right\rangle _{ABC}),
\end{equation}
where the notion and the states $\left| \Phi _{0}\right\rangle $ and $\left|
\Phi _{1}\right\rangle $ are defined as before. In order to create a program
state corresponding to the operation $U^{^{\prime }}$, the state $\left|
\Phi \right\rangle $ is subjected to a transformation $U_{P}^{^{\prime
}}\otimes I_{ABC}$: 
\[
\left| P_{U^{^{\prime }}}\right\rangle =U_{P}^{^{\prime }}\otimes
I_{ABC}\left| \Phi \right\rangle =\frac{1}{\sqrt{2}}(e^{i\frac{\beta }{2}%
}\left| 0\right\rangle \left| \Phi _{0}\right\rangle +e^{-i\frac{\beta }{2}%
}\left| 1\right\rangle \left| \Phi _{1}\right\rangle ), 
\]
where only the port qubit is transformed by the operation $U^{^{\prime }}$.
The quantum gate array firstly flips the data qubit by Pauli operator $%
\sigma _{x}$, then makes joint Bell-state measurement M on data qubit D and
port qubit P, and outputs four results. In detail, here is how it works: 
\begin{eqnarray}
\sigma _{x}\left| d\right\rangle \otimes \left| P_{U^{^{\prime
}}}\right\rangle &=&(a\left| 1\right\rangle +b\left| 0\right\rangle )\otimes 
\frac{1}{\sqrt{2}}(e^{i\frac{\beta }{2}}\left| 0\right\rangle \left| \Phi
_{0}\right\rangle +e^{-i\frac{\beta }{2}}\left| 1\right\rangle \left| \Phi
_{1}\right\rangle ) \\
&=&\frac{1}{2}[be^{i\frac{\beta }{2}}(\left| \Phi ^{+}\right\rangle +\left|
\Phi ^{-}\right\rangle )\left| \Phi _{0}\right\rangle +ae^{-i\frac{\beta }{2}%
}(\left| \Phi ^{+}\right\rangle -\left| \Phi ^{-}\right\rangle )\left| \Phi
_{1}\right\rangle  \nonumber \\
&&+be^{-i\frac{\beta }{2}}(\left| \Psi ^{+}\right\rangle +\left| \Psi
^{-}\right\rangle )\left| \Phi _{1}\right\rangle +ae^{i\frac{\beta }{2}%
}(\left| \Psi ^{+}\right\rangle -\left| \Psi ^{-}\right\rangle )\left| \Phi
_{0}\right\rangle ]  \nonumber \\
&=&\frac{1}{2}[\left| \Phi ^{+}\right\rangle (be^{i\frac{\beta }{2}}\left|
\Phi _{0}\right\rangle +ae^{-i\frac{\beta }{2}}\left| \Phi _{1}\right\rangle
)+\left| \Phi ^{-}\right\rangle (be^{i\frac{\beta }{2}}\left| \Phi
_{0}\right\rangle -ae^{-i\frac{\beta }{2}}\left| \Phi _{1}\right\rangle ) 
\nonumber \\
&&+\left| \Psi ^{+}\right\rangle (be^{-i\frac{\beta }{2}}\left| \Phi
_{1}\right\rangle +ae^{i\frac{\beta }{2}}\left| \Phi _{0}\right\rangle
)+\left| \Psi ^{-}\right\rangle (be^{-i\frac{\beta }{2}}\left| \Phi
_{1}\right\rangle -ae^{i\frac{\beta }{2}}\left| \Phi _{0}\right\rangle )]. 
\nonumber
\end{eqnarray}
As has been discussed above, when the measurement result from M gives
eigenvalues corresponding to $\left| \Phi ^{+}\right\rangle $ or $\left|
\Phi ^{-}\right\rangle $, the three data qubits are projected onto or can be
effectively converted by Pauli operator $\sigma _{z}$ into a state 
\begin{equation}
\left| X^{^{\prime }}\right\rangle =(be^{i\frac{\beta }{2}}\left| \Phi
_{0}\right\rangle +ae^{-i\frac{\beta }{2}}\left| \Phi _{1}\right\rangle ).
\end{equation}
Similarly, dropping the ancillary qubit and tracing out qubit C or B, we
have 
\begin{eqnarray}
\rho _{_{B}} &=&Tr_{A,C}(\left| X^{^{\prime }}\right\rangle \left\langle
X^{^{\prime }}\right| )=\rho _{_{C}}=Tr_{A,B}(\left| X^{^{\prime
}}\right\rangle \left\langle X^{^{\prime }}\right| ) \\
&=&\frac{5\left| b\right| ^{2}+\left| a\right| ^{2}}{6}\left| 0\right\rangle
\left\langle 0\right| +\frac{5\left| a\right| ^{2}+\left| b\right| ^{2}}{6}%
\left| 1\right\rangle \left\langle 1\right|  \nonumber \\
&&+\frac{2}{3}a^{*}be^{i\beta }\left| 0\right\rangle \left\langle 1\right| +%
\frac{2}{3}ab^{*}e^{-i\beta }\left| 1\right\rangle \left\langle 0\right| . 
\nonumber
\end{eqnarray}
The correctly transformed version of input data state we desired is $%
U^{^{\prime }}\left| d\right\rangle =be^{i\frac{\beta }{2}}\left|
0\right\rangle +ae^{-i\frac{\beta }{2}}\left| 1\right\rangle $. According to
the Eq.(9) we get the fidelities of the states $\rho _{B}$ and $\rho _{C}$
as 
\begin{equation}
F_{B}^{^{\prime }}=F_{C}^{^{\prime }}=\frac{5}{6}.
\end{equation}
For other two results of measurement M, the quantum gate array yields ruined
state from which the state $U^{^{\prime }}\left| d\right\rangle $ we desired
can not be retrieved. Therefore for the operation anti-commuting with Pauli
operator $\sigma _{z}$ the modified quantum processor yields two approximate
transforms with successful probability of $\frac{1}{2}$. The fidelity of
each successful data output is $\frac{5}{6}$. In the same way, the results
of measurement give the information about whether the process is successful.

We have designed a two-output programmable processor respectively for sets
of operations on single qubit commuting and anti-commuting with Pauli
operator $\sigma _{z}$. The programmable processor implements the restricted
operations on arbitrary data input state with successful probability of $%
\frac{1}{2}$ and provide two outputs with fidelity of $\frac{5}{6}$. It is
noticed that the states $\rho _{_{B}}$ and $\rho _{_{C}}$ just are mixed
states since $Tr((\rho _{_{B}})^{2})=Tr((\rho _{_{C}})^{2})=\frac{13}{18}<1$%
. That is, we apparently do not have complete knowledge about $U\left|
d\right\rangle $ (or $U^{^{\prime }}\left| d\right\rangle $).

\section{Discussion}

We have presented how to construct a two-output programmable processor. It
is necessary to point out that the programmable quantum processor is
approximate and probabilistic. The set of operations which it applies on the
input data state is limited. It comes from three restrictions. One is that
in itself no universal quantum gate array exists in deterministic fashion.
The other is that it involves a telecloning procedure complying with
no-cloning theorem \cite{w.k}. The third lies in, during the creation of
program state, executing the operation on the port qubit which later is
subjected to the joint measurement with the data qubit in the Bell basis.
Simple algebra shows that $(I\otimes U)(|00>+|11>)=(U^{T}\otimes
I)(|00>+|11>)$, where $U^{T}$ denotes the transpose of $U$ with respect to
the computational basis. Unlike the quantum gate array proposed by Nielsen
and Chuang \cite{nielsen and chuang}, we has to concern transforming $%
U^{T}\left| d\right\rangle $ to $U\left| d\right\rangle $. It is easily
proved that for an unknown operation $U$ and an arbitrary state $\left|
d\right\rangle $, the transformation can not be completed effectively.
Nielsen-Chuang processor \cite{nielsen and chuang} produces one data output
for arbitrary operation on the input data with successful probability of $%
\frac{1}{4}$ and the perfect fidelity of 1, and Vidal's \cite{vidal and
cirac} one-to-one processor can be programmed to implement restricted set of
operations on input data with the successful probability of $p\geq \frac{1}{2%
}$ and the same perfect fidelity. Compared with them our processor creates
two data outputs for restricted set of operations on input in price of
decreasing of the fidelity and probability of success. But our processor is
more powerful in remote control and distributed computation. For example, if
we expect to rotate two spin-$\frac{1}{2}$ particles in different states
simultaneously, we can use our two-output programmable quantum processor as
a program encoder and match one particle and one data output of the
processor respectively, then feed them into another quantum gate arrays such
as the Nielsen-Chuang \cite{nielsen and chuang} and Vidal's \cite{vidal and
cirac}.

On the other hand, a sequential scheme, in which a Nielsen-Chuang processor
on teleportation \cite{nielsen and chuang} is followed by a 1$\rightarrow $2
telecloning procedure, may achieve the same purpose as our two-output
processor. However, the sequential scheme accomplishes arbitrary
transformation on input data state with less probability of success and the
same fidelities compared with the two-output processor. Furthermore, the
sequential scheme is more complex and more consumptive. It not only consumes
more1 ebit of entanglement and 2 bits of classical information communication
than the two-output processor, but also involves more physical qubits and
quantum operations than the two-output processor. In a sense, our two-output
processor reduces the resource consumption and complexity scales of the
information process. If a general approximate 1$\rightarrow $2 cloning
procedure follows the Nielsen-Chuang processor instead of 1$\rightarrow $2
telecloning procedure in the sequential scheme, besides the above
disadvantages, whether the two data output registers of the sequential
scheme can be separate from each other is to be queried. Since our
two-output processor takes advantage of the nonlocal cloning of telecloning
procedure and has two distant outputs, it is suitable for remote control and
distributed computation of multiple spatial location. At this point our
two-output processor is more useful than the sequential scheme.

The two-output programmable quantum processor can be straightforward
generalized to the case of multi-output and multi-qubit operation. It is
easy to understand that the fidelity decreases with the increase of the
numbers of data output and the probability decreases exponentially with the
number of qubits the operation involved. Whether there is more effective
scheme of multi-output quantum processor is still an open question.

\section{Conclusion}

In this paper, we have proposed a multi-output programmable quantum
processor. As a result of three restrictions, our processor is approximate
and probabilistic, and the set of operations which it can implement is
restricted. As an example, we give a two-output programmable quantum
processor for sets of operations commuting and anti-commuting with Pauli
operator $\sigma _{z}$, which is based on telecloning procedure. With
probability of $\frac{1}{2}$, it yields two approximately transformed data
outputs in the fidelity of $\frac{5}{6}$, which can be distributed to next
processors for different task, respectively. And due to the property of
entanglement, the data registers in terminal can separately be located, even
far from the processor if it allows LOCC (local operation and classical
communication). Therefore the multi-output processor can be extended to
distributed computation and remote control of multi-user. Compared with a
sequential scheme in which a Nielsen-Chuang processor is followed by an
approximate cloning procedure, our processor reduces the resource
consumption and complexity scales of the information process and has two
distant outputs. But the probability of success and the fidelity of data
output of the multi-output processor decrease with the numbers of qubits
involved in the computation and of data outputs required. It would be
extremely intriguing to know if it is possible to build a more effective
multi-output processor.

\begin{center}
{\bf Acknowledgments}
\end{center}

This work has been financially supported by the National Natural Science
Foundation of China under the Grant No.10074072.


\begin{references}
\bibitem{nielsen and chuang}  M. A. Nielsen and I. L. Chuang, Phys. Rev.
Lett. 79, 321 (1997).

\bibitem{vidal and cirac}  G. Vidal, L. Masanes and J. I. Cirac, Phys. Rev.
Lett. 88, 047905 (2002).

\bibitem{hillery and ziman}  M. Hillery, V. Buzek and M. Ziman, Phys. Rev. A
65, 022301 (2002).

\bibitem{hillery and buzek}  M. Hillery, M. Ziman and V. Buzek, Los Alamos
e-print arXiv quant-ph/0205050 (2002).

\bibitem{telecloning}  M. Murao, D. Jonathan, M. B. Plenio, and V. Vedral,
Phys. Rev. A 59,156(1999).

\bibitem{1022301}  E. F. Galvao and L. Hardy, Phys. Rev. A 62,022301(2000).

\bibitem{w.k}  W. K. Wootters and W. H. Zurek, Nature (London) 299,
802(1982).
\end{references}
\end{document}